# Electrostatic Force Suspended-Air Multilayer (EPAM) Structure for Highly Transparent Energy Efficient Windows

Ying Zhong,[*ab] Rui Kou,[b] Renkun Chen[c] and Yu Qiao[b]

a. Department of Mechanical Engineering, University of South Florida, 4202 E. Fowler Avenue, Tampa, FL 33620, USA. Email: yingzhong@usf.edu

b. Department of Structural Engineering, University of California at San Diego, 9500 Gilman Drive, La Jolla, CA 92093, USA.

c. Department of Mechanical Engineering, University of California at San Diego, 9500 Gilman Drive, La Jolla, CA 92093, USA.

† Y Zhong and R Kou Contribute equally to this work.

**Abstract**: The poor thermal insulating building windows, especially single pane windows, are wasting ~7% of the total energy consumed by the U.S. every year. This paper presents an electrostatic force suspended polymer-air multilayer (EPAM) structure as a highly transparent and energy efficient window retrofitting solution for low-income single pane window users. To provide controllable and stable suspension force between large size compliant polymer films without deteriorating their visual transmittance (Vt), corona discharge (CD) was induced to permanently charge polymer films, leading to controllable permanent surface potential difference between two sides of CD charged films. Liquid-solid contact electrification (CE) was combined with CD to realize precise control of the surface potential of each side of polymer films. CD+CE treated films can obtain programmable and stable electrostatic forces to form resilient EPAM structures for cost and energy effective window retrofitting purpose. The resilient EPAM retrofitted windows can provide U-factor lower than 0.5 Btu·hr$^{-1}$·ft$^{-2}$·°F$^{-1}$, Vt higher than 70%, haze lower than 1.6% at cost lower than \$7.4/ft$^2$, at least 10 times cheaper than double pane windows. The energy saved by EPAM can reach as much as the order of $10^6$ kJ/m$^2$ per year. The CD+CE surface potential programming solution also provides a highly repeatable and controllable way for other electrical potential related technologies.

**Keywords**: Window energy; Polymer electrification; U-factor; Visual transimttance

# 1. Introduction

According to the Buildings Energy Data Book, the U.S. buildings consume 38~40 quads (1 quad=$10^{15}$ BTU=$1.055 \times 10^{18}$ J) of energy per year,[1] nearly 40% of the total energy consumption in the U.S.[2–4] Among which, ~6.9 quads of energy is lost and wasted by building windows because of their poor thermal insulating property. The estimated overall average U-factor of all windows is as high as 0.75 Btu·hr$^{-1}$·ft$^{-2}$·°F$^{-1}$.[5–8] Among them, the 30~40% single-pane windows with U-factor of ~1.2 Btu·hr$^{-1}$·ft$^{-2}$·°F$^{-1}$ are consuming the major portion of the energy.[2,5,6] If they can be replaced by Energy Star windows with U-factor of 0.3~0.5 Btu·hr$^{-1}$·ft$^{-2}$·°F$^{-1}$, there will be ~3.5 quads of energy saved per year in the U.S., which is ~10% of the total energy consumed by all buildings, and save ~$45 billion energy cost. [2,5,6]

However, replacing or upgrading single-pane windows with double-pane windows are expensive and slow, with expense of $800~$1,500 per square meter, approximately up to $20,000 per household.[9,10] Fully replacing single-pane windows may take the U.S. as long as four decades, leading to extra energy waste of >120 quads.[2,4,5] Compared with window replacing, window retrofitting is easier, faster, more economical, and less labour intensive.[11–16] Currently used retrofitting methods include storm windows, attaching low-emissivity (low-e) films and thermal insulating materials. Storm windows need to add another framed pane to the original windows, but they are heavy, and windows cannot be conveniently opened or closed.[17,18] Attaching low-e films on the original single panes offers a relatively cost-efficient method to reduce radiative heat transfer across widows in the infrared (IR) domain.[19–21] However, the poor visual transmittance (Vt) and colour rendering index (CRI) are the largest hurdles for wide application on the market; plus, water condensation deteriorates low-e effect significantly, making it use-less for cold weather regions.[22–25] Attaching thermal insulating materials is another choice, but it requires the material to be highly transparent with superior thermal insulating properties. Aerogel has been a promising candidate with ultralow-thermal conductivity as low as 0.013 W/(m·K) and transmittance as high as 99%.[26–29] However, the haze of most aerogel is over 5%, which is unfavoured for window applications.[30] It is expensive and difficult for mass production. It is also challenging to overcome their brittleness and super-low flexural strength without deteriorating the transmittance.[31] The ideal window retrofitting technique should provide the lowest U-factor at the lowest economical, Vt and maintenance cost with long-term mechanically resilience.

This paper presents a novel concept of window retrofitting by applying electric force suspended polymer-air alternating multi-layer (EPAM) structures onto single-pane windows to create highly transparent and energy efficient windows (window+). EPAM (3~5 mm in total thickness) is constructed by assembling a few thin transparent polymer films (25~50 μm) in parallel. The thin and compliant polymer films are suspended stably with air gaps by the invisible electrostatic repulsion force created by controllable amount of electrostatic charges created on the films through our newly developed surface potential control technology.

EPAM provides an ideal solution to reduce the U-factor for windows in all the three thermal transfer paths: 1) conduction: >93% of the structure is made by air with thermal conductivity $\kappa_{air}$ as low as 0.024 W·m$^{-1}$·K$^{-1}$, meaning the overall thermal conductivity $\kappa$ can be lower than 0.03 W·m$^{-1}$·K$^{-1}$, which is low enough to provide good thermal insulating for the window+;[32,33] 2) radiation: each layer of film can block a portion of thermal radiation leading to a low emissivity in the range of 0.1~0.2, which is equivalent to the best low-e films on the market;[34–37] 3) convection: with air gap reduced to narrower than 1 mm, the convection can be reduced to less than 2% of regular double-pane windows, reducing the effective thermal conductivity.[38–40] Overall, through blocking all three thermal transfer paths, the U-factor of the single-pane window can be effectively reduced to less than 0.5 Btu·hr$^{-1}$·ft$^{-2}$·°F$^{-1}$ by retrofitting them with the cost effective EPAM structures.

Collapsed structures lead to reduction in Vt and increase in U-factor. Therefore, to guarantee the effectiveness of EPAM, the air gap thickness between each layer in the multilayer structure has to be stably maintained and controlled. However, the windows are as large as meter size, and polymer films with thickness of tens of micrometres are too compliant to keep the air gaps stable.[41] Conventional ways stabilize the air gap between compliant films include adding physical supporters or compositing the films to create magnetic repulsion forces. However, both ways deteriorate the Vt of the multiple layers. In this paper, we innovatively propose to utilize electrostatic force to stably and uniformly separate polymer films at long-term without deteriorating the Vt.[42] To maintain durable and controllable air gap, corona discharging (CD) was utilized to create "permanent" charges in transparent polymer films without deteriorating their Vt.[43–48] CD is a broadly used technology for polymer film surface modification in industry, suitable for mass production with treatment speed up to 500 m/min.[49] However, it was discovered that the surface potential fluctuates significantly among samples treated with the same parameters, which cannot contribute to controllable the repulsion force between films.[43] To address this issue, better understanding of the charging behaviour of CD is required, and precise control of the surface potential of polymer films has to be realized. Herein, the surface potential behaviour of CD charged polymer films were systematically studied, and a new path to control the surface charge density of polymer films by combining CD and liquid-solid contact electrification (CE) was discovered and investigated. Based on which, the precise control of the electrical interaction force between films has been realized. The stability of the surface potential and the resilience of the EPAM structure was confirmed. And the U-factor of the EPAM retrofitted windows+ was tested.

## 2. Experimental

### 2.1 Materials

Considering cost, Vt, chemical and UV resilience, and charge stability, films used in this test are the clear moisture-resistant polyethylene terephthalate (PET) films obtained from McMaster-Carr (Product No. 8567K14) with thickness of 125 μm. The price on McMaster-Carr is only $44.67 for 25' x 40", meaning the material cost to retrofit windows with 4-layer EPAM is as low as $2.2 for one square foot at retail price, over 50 times lower than double-pane window replacing.

### 2.2 Corona Discharging (CD)

A corona charging system (Fig. 1a,b) was set up to electrify selected polymer films. The setup consisted of a discharging needle electrode, a grid, a grounded electrode, and two high voltage power supplies. A sharp tungsten needle of 0.75-inch-long and 0.059 inch in diameter was employed as the discharge needle electrode. The diameter of the needle tip was ~0.1 mm. The discharge electrode can be a tungsten wire as well to charge larger area. The stainless steel 304 grid was placed between the needle and the sample, with the wire diameter of 0.016 inch and the mesh size #20. Polymer film samples, usually 6 inch × 6 inch large, were placed on the grounded electrode, a mirror-polished stainless-steel plate. The distance between the grounded electrode with the grid is 1 cm, and between the needle is 4 cm. PET films were firstly ultrasonically cleaned in isopropyl alcohol (IPA) for 5 min and then in de-ionized (DI) water for another 5 min, then blow dried. The needle voltage and the grid voltage were controlled by two polarity switchable Glassman Co. Lt., FJ Series 120 Watt regulated high voltage DC power supplies, respectively, with the same polarity. The voltage could be adjusted in the range from 0 to ±40 kV. The voltage of the needle ($V_n$) was 10 kV, and the voltage on the grid ($V_g$) ranges from 1 kV to 3 kV.

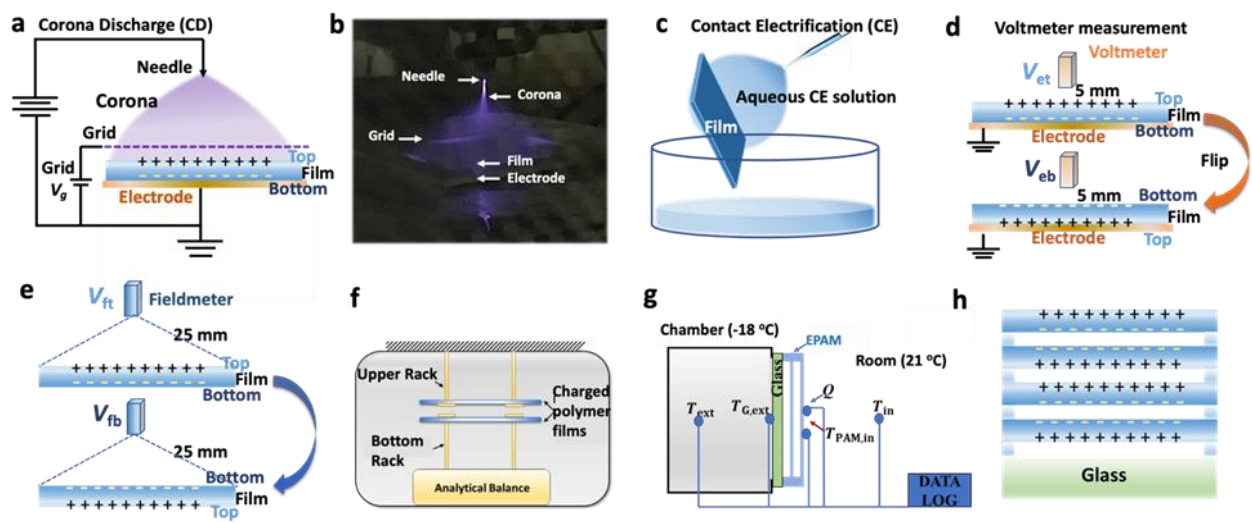

**Fig. 1** Experiment setups. a) Schematic of the corona discharge (CD) system, b) optical photograph of corona discharge treatment on polymer film, c) schematic of contact electrification (CE) treatment on polymer films, d) schematic of measuring the surface potential of the charged films with voltmeter, the not being tested side is grounded, and the potential on the other side is measured by flipping the film over, e) schematic of measuring surface potential with fieldmeter while the film is suspended (no electrode in contact with the film), and both sides were separately measured, f) schematic of the electrostatic interaction force measurement setup, g) schematic of U-factor measurement set up, h) EPAM structure schematic.

**2.3 Contact Electrification (CE)**

Contact electrification (CE) experiments were carried out by a variety of liquids. To guarantee the low cost and accessibility of the treatment liquids, aqueous solution of isopropyl alcohol (IPA) and NaCl solutions were selected in this application. The IPA solutions were prepared by ultrasonically mixing IPA with DI water for 3 min. The IPA concentration, $C_{IPA}$, was 0, 5, 10, 20, 30, 50, or 100 vol.%. Aqueous solutions of sodium chloride (NaCl) were prepared by ultrasonically dissolving NaCl in DI water for 5 min; the NaCl concentration, $C_{NaCl}$, was 0, 0.5, 1, 2, 5, or 10 wt.%. For the CE treatment, the polymer films were vertically immersed in a selected liquid for 3 sec and pulled out with velocity of ~2.5 cm/s. For the CD+CE treatment, the films were firstly treated with CD, then one side of the corona charged film was treated by the NaCl or IPA solution, as the liquid flew through the film surface with a constant rate around 1 mL/s (Fig. 1c). PET films CD charged with $V_g$=-1, -2, and -3 kV were treated with above solutions. At least 5 samples were tested and characterized for each parameter setting.

## 2.4 Surface Potential Measurement

The charged sample was characterized with voltmeter (Trek, Model 344 electrostatic voltmeter) and fieldmeter (Simco, FMX-004 electrostatic fieldmeter). For the voltmeter measurement (Fig. 1d), the polymer films were placed on the grounded electrode, and a Kelvin vibrating probe was positioned 5 mm away from the sample. It could measure the average surface voltage over a small surface area less than 10 mm in diameter. Since the film was backed by the grounded electrode, the voltmeter reading ($V_v$) reflected the voltage distribution of the side facing the probe. To map the surface potential distribution, $V_v$ was measured every 0.5'' along the diagonal of each sample as the surface voltage distribution is symmetric. After the top side of the film (the side facing the needle and grid during charging) was measured to obtain $V_{vt}$, the sample was flipped over and the voltage along its bottom side ($V_{vb}$) was measured, with the top side firmly attached to the grounded electrode. For the fieldmeter measurement (Fig. 1e), the probe was placed 25 mm away from the polymer film, and the film was suspended in air. Any contact with other objects was carefully avoided. The fieldmeter reading ($V_f$) reflected the overall surface potential contributed by all charges from both sides of the film. It directly determines the electrostatic force between films. All the samples were preserved at ambient temperature in an environmental chamber with the relative humidity (RH) controlled at 45%. Their surface potential on both sides were monitored for at least 7 days.

## 2.5 Electric Force Measurement and Modelling

A measurement system (Fig. 1f) was set up to characterize the electrostatic force between electrified polymer films. It consists an analytical balance (Ohaus Explorer Scale-220 grams with resolution of 0.0001 grams), two polycarbonate (PC) racks as sample holders, and a glass shielding cage. The bottom rack was fixed on the sample plate of the balance. The upper rack was affixed on the ceiling of the cage. One electrified film was first placed on the bottom rack, and the scale was calibrated to zero. Another film was then placed on the upper rack, with the edges aligned with the bottom film. It approaches to the bottom film with controllable distance between them. The samples were carefully handled such that their surfaces were not in contact with any solid or liquid objects. The repulsive force was obtained through the reading of the balance.

FEA modelling was performed to analyse the electrical repulsive pressure between two 125 $\mu$m-thick, 76 mm-large free-standing electrified PET films. COMOSL Multiphysics Electrostatic Module was employed. The dipolar charged density applied on the films was 0.67 mC/m$^2$, 0.45 mC/m$^2$ and 0.22 mC/m$^2$, representing surface potential of 1 kV, 2 kV and 3 kV for 3 inch$^2$ large sample. The dielectric constant of PET film and air in between was set to 3 and 1, respectively. Free Tetrahedral mesh was employed.

**2.6 EPAM Structural Resilience Characterization**

The structural resilience of EPAM structures was characterized through repeated compression tests. The EPAM samples was 0.3 m in length and width, and their initial thickness was 3 mm. The PET films were 0.25 mm thick and the air gap in between was 1.25 mm. A Type 5582 Instron machine was employed to control a loading rod which had a 5-mm-radius round tip. It was compressed by the Instron machine with a speed of 0.01 mm/s at the centre of the EPAM sample for 2000 times. The compression displacement was 2.5 mm along the EPAM thickness direction.

**2.7 U-factor Measurement**

U-factor reflects the overall thermal transmittance, including conduction, convection, and radiation. Lower U-factors lead to better thermal insulation properties and higher energy efficiency. A *U*-factor measurement system was designed and fabricated following ASTM C1199[50], ASTM C518,[51] NFRC-100,[52] and NFRC-102.[53] Figure 1g is the schematic of the U-factor measurement setup. The environmental chamber's size is 0.39-m long, 0.47-m wide and 0.47-m tall. Cold side temperature in the chamber was maintained at -18 °C ($T_{ext}$), mimicking the cold external environment. A EPAM sample was installed on the glass wall of the chamber. The EPAM sample was attached well onto the glass, facing the lab environment, representing the inside of the warmer room in cold weather ($T_{in}$). $T_{ext}$, $T_{in}$, the exterior surface temperature of glass ($T_{G,ext}$), and the interior surface temperature of EPAM ($T_{EPAM,in}$) were monitored by four Omega type-K thermocouples. A gSKIN-XI heat flux sensor was installed on the interior surface of the EPAM to monitor the heat flux.

According to ASTM C1199 and NFRC-100, U-factor is calculated as the reciprocal of the total thermal resistance

$$U = \left(\frac{1}{h_{\text{ST,ext}}} + \frac{T_{\text{EPAM,in}} - T_{\text{EPAM,ext}}}{q} + \frac{1}{h_{\text{ST,in}}}\right)^{-1} \qquad (1)$$

where, the standard boundary heat transfer coefficients to the exterior side ($h_{\text{ST,ext}}$) was set to 30 W/(m²·K); the interior side heat transfer coefficient ($h_{\text{ST,in}}$) was calculated according to NFRC-100 as[54]

$$h_{\text{ST,in}} = 1.46 \left[\frac{T_{\text{in}} - T_{\text{EPAM,in}}}{H}\right]^{0.25} + \sigma\varepsilon\left[\frac{(T_{\text{in}} + 273.16)^4 - (T_{\text{EPAM,in}} + 273.16)^4}{T_{\text{in}} - T_{\text{EPAM,in}}}\right]; \qquad (2)$$

where, $H$ is the sample height (0.3 m); $\sigma = 5.67 \times 10^{-8}$ W/m² · K⁴ is the Stefan Boltzmann constant; and $\varepsilon = 0.14$ is the surface emissivity of low-e coating.

## 2.8 Optical Property Characterization

Visual transmittance, haze, and CRI of EPAM were characterized in the spectral range from 380 nm to 750 nm by a JASCO V770 UV-VIS spectrometer, following ASTM D1003.[55] The spectrophotometer was equipped with an integrating sphere accessory. During the measurement, a EPAM sample was placed in front of the sphere entrance port, through which light can pass into the sphere and be fully collected. The visible light transmittance ($V_t$) is defined in ASTM G173 as $V_t = \frac{\int \tau(\lambda) E(\lambda) d\lambda}{\int E(\lambda) d\lambda}$ where, $E(\lambda)$ is the solar spectral irradiance, $\tau(\lambda) = \tau_t / \tau_i$, $\tau_t$ is the light transmitted through the sample, and $\tau_i$ is the incident light. Haze can be calculated from the sample diffusion $\tau_{s,d}$ and the instrument diffusion $\tau_{i,d}$ [56] Haze= $\tau_{s,d}/\tau_t - \tau_{i,d}/\tau_i$.

The transparency colour perceptions are designed to represent the colours perceptible to human eyes. Based on the measured visual transmittance spectrum of EPAM, the CIE chromaticity values and the colour rendering index can be obtained based on the previous studies [56–60].

## 3 Results and Discussion

### 3.1 Corona Discharge (CD)

Corona discharge (CD) was conducted on PET films as described in the experimental session. As mapped in Fig. 2a, about 50 cm² of the central part of the films was electrified to the voltage the same as the grid (2 kV here) within 10 s. The uniformly charged area increased to ~100 cm² if the CD processing time was increased to 20 s. Increasing the time further to 180 s did not increase the effective area significantly. As tested in Fig. 2b varying the grid voltage ($V_g$) can lead to the same change of the surface potential of the films, indicating the voltage on the electrified film can be easily controlled by $V_g$. By measuring the potential of both sides of the film with the voltmeter, it was discovered that the voltages on the top ($V_{vt}$) and bottom side ($V_{vb}$) of the film shared the same absolute value but opposite polarities, with the top side showing the same polarity with the discharge electrode. As seen in Fig. 2c, the charged area can be easily increased to the entire film by replacing the needle discharge electrode with a wire electrode with a length not less than the film size. The voltage distribution on the entire surface of the film was uniform and also equal to the voltage of the grid voltage. Further increasing the wire length can allow us to charge films with larger area with similar length as the wire length. Above studies indicate that CD can treat large-size films within a short time of 20 s, while providing the capability to control the surface potential. The stability of the surface potential was tested for as long as 1 months, and the decay was less than 20% as shown in Fig. 2d. The decay of the charges tends to slow down with time. It indicates that the charges formed by CD are stable enough to separate the films in the EPAM structure for at least several years.

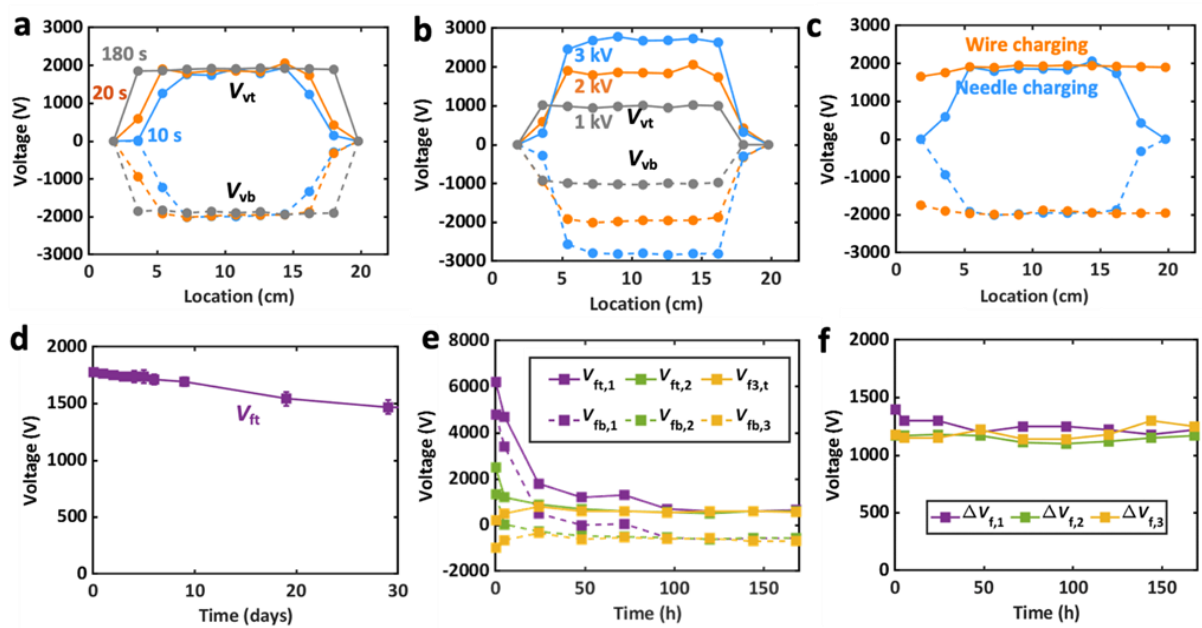

**Fig. 2** Surface potential of PET films after corona discharge (CD) treatment. a) Voltmeter measured voltage distribution along the diagonal line of the PET film at the top ($V_{vt}$) and bottom ($V_{vb}$) sides charged after different times of 10 s, 20 s, and 180 s, b) $V_{vt}$ and $V_{vb}$ charged at different grid voltages of 1 kV, 2 kV and 3 kV, c) $V_{vt}$ and $V_{vb}$ distribution of PET films charged with tungsten wires vs. with needles, d) surface voltage decay for 1 month, e) $V_{ft}$ and $V_{fb}$ of three different PET films charged with the same CD parameters, f) $\Delta V$ ($\Delta V = V_{ft} - V_{fb}$) *change with time* of three different PET films charged with the same CD parameters.

Measurement of the surface potential while the films were suspended was conducted by the fieldmeter. It detects the overall surface potential of the films created by all charges on both sides, which directly determines the interaction between films. Figure 2e is the fieldmeter measured data ($V_f$) of different pieces of PET films charged with the same CD parameters. The interesting phenomenon discovered here is that even though three PET films were treated with the same CD parameters, the fieldmeter measured data were extremely different. Here, $V_{ft}$ and $V_{fb}$ represents the surface potential of the top and bottom side, respectively. However, the voltage difference ($\Delta V=V_{ft}-V_{fb}$) were almost the same for every tested sample (Fig. 2f).

The mechanism of this phenomenon has been elaborated in our previous publication.[43] There are two types of charges contributing to the surface potential, which are the major dipolar charges created in the films by CD and free charges caused by contact electrification (CE) whose density is only less than 0.1% of the injected charges. The major CD charges controls $\Delta V$; and the minor CE charges controls the absolute values of $V_{ft}$ or $V_{fb}$. If one film is only charged by CD, only $\Delta V$ can be controlled, but not $V_{ft}$ or $V_{fb}$ who directly determine the interaction force between films in EPAM. Thus, even though CD can create long-lasting charges, to obtain structurally stable EPAM, better ways to control $V_{ft}$ and $V_{fb}$ is required.

**3.2 Contact Electrification**

As mentioned above, CE created surface charges are the key factors determining the value of $V_{ft}$ and $V_{fb}$. CE is a behaviour which researchers have not obtained good control or understanding. It has been proved that the capability to obtain controllable surface charges by solid-slid contact electrification (CE) is poor, because of the challenges in controlling contact area or force.[61] Promisingly, we discovered that based on double layer theory (Fig. 3a) liquid-solid CE (LS-CE) can provide good controllability over $V_{ft}$ and $V_{fb}$, as the ingredient of the liquid and the flow speed/contact force can be easily adjusted. As this application is targeted at window retrofitting, the most commonly used environmentally friendly solutes are recommended to reduce the cost and provide the widest access. Thus, we utilized IPA and NaCl solutions to control the free charges induced surface potential of polymer films.

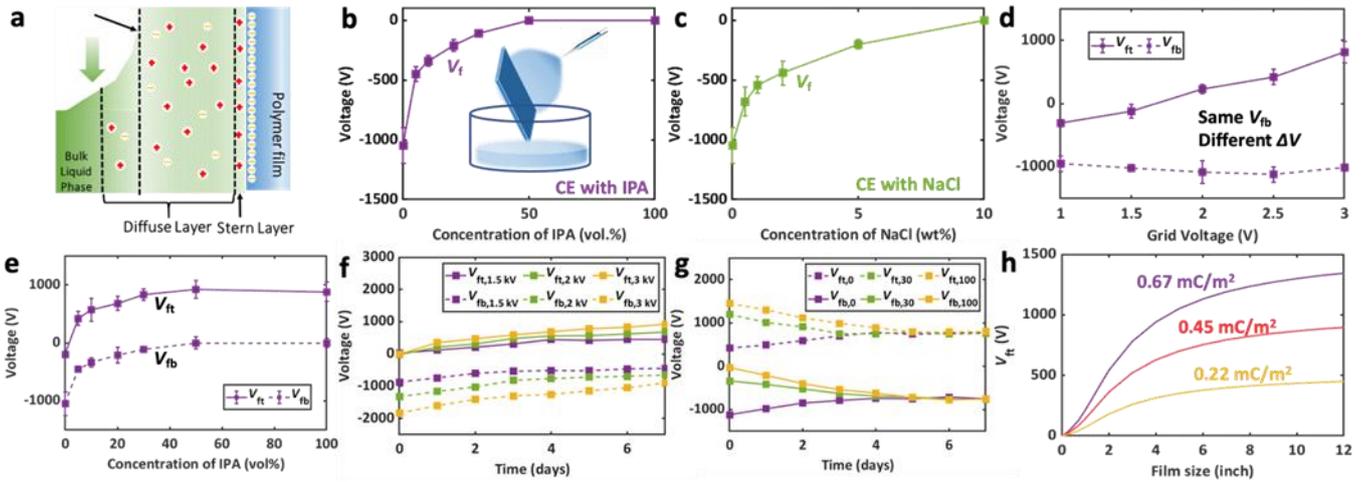

**Fig. 3** Surface potential of PET films after liquid-solid contact electrification (LS-CE) treatment. a) Schematic of the double layer theory, b) CE treated with IPA aqueous solution with concentration ranging from 0 to 100%, c) PET films CE treated with NaCl aqueous solution with concentration ranging from 0 to 10%, d) surface potential of PET films ($V_{ft}$ and $V_{fb}$) obtained after firstly charged with different CD induced $\Delta V$ ranging from 1 to 3 kV, then bottom side CE treated with DI water, e) surface potential of PET films firstly charged with CD induced $\Delta V$ of 1 kV, then bottom side CE treated with IPA solution of different concentrations, f) surface potential evolution for 7 days after CD+CE treatment, CD induced $\Delta V$ of 1 kV, 1.5 kV, and 2 kV, CE induced $V_{ft}$=0 by pure IPA, g) surface potential evolution for 7 days starting from different $V_{ft}$ and $V_{fb}$, but the same $\Delta V$ =2 kV, h) surface potential of charged film increases with the size of the film and charge density.

As plotted in Fig. 3b, if raw PET films are CE treated with pure DI water, they can obtain surface potential of ~-1 kV. If DI water is mixed with IPA or NaCl, the obtained surface voltage decreases with the increase of their concentration. Specifically, when the PET films were treated with 5 vol.% IPA solution, the surface voltage was about -0.4 kV, but if treated with 100% IPA, the surface potential was reduced to 0. In Fig. 3c, the obtained surface potentials were -0.7 kV and 0 for NaCl concentrations of 5 wt.% and 10 wt.%, respectively. The measured surface potential data from different times of treatment was highly consistent, indicating the good controllability of LS-CE.

The double layer theory can be utilized to explain this phenomenon.[62] As schematized in Fig. 3a, upon contact with liquid solution, the surface groups of the polymer film will be ionized or dissociated, rendering the surface negatively charged. The liquid-induced charge density is denoted as $\sigma_0$. The electric double layer (EDL), including a stern layer and a diffusive layer, will be created at the surface the film due to electrostatic attraction. When the liquid flows across the surface, the liquid layer within the shear plane in the EDL will stay on the polymer surface, and the charge density of the remnant ions within the shear plane can be calculated by [63] $\sigma_f = \sigma_0 e^{-z_s\sqrt{(e^2/\varepsilon k_B T)\sum_i 1000 z_i^2 m_i N_A}}$ where, $z_s$ is the shear plane distance, $\varepsilon$ is the liquid permittivity, $k_B$ is the Boltzmann constant, $T$ is temperature, $z_i$ is the ion charge, $m_i$ represents the molarity of ion specie $i$, and $N_A$ is Avogadro's number. The $\sigma_f$ reduces with the salt concentration ($m$), matching with the results in Fig. 3b and 3c. Other solutions can also be utilized to create controllable surface potential on polymer films through liquid CE treatment, and the obtained surface potential. For example, if the side group length of the solute is increased, the $z_s$ will be increased which also increase the $\sigma_f$ and $V_f$.[62]

Thus, by controlling, the ingredient and concentration of the liquid used in the CE treatment is a highly efficient way to control $V_{ft}$ and $V_{fb}$. However, both sides of the film share the same surface potential if only treated with CE. And the potential decays back to 0 within hours and even minutes, as the charge density is very low, and charges distribute only at the surface of the films. Even though CE can precisely control the $V_{ft}$ and $V_{fb}$, it cannot provide long-term separation for films in EPAM.

**3.3 CD+CE Surface Potential Control**

It has been thoroughly discussed above that CD can create long-lasting surface potential difference between two sides of the films, but not able to control the surface potential on each side. While CE can control the value of surface potential, but does not have the capability of long-lasting. Therefore, in this paper, we propose to combine CD and CE together to take advantage their benefits and overcome their draw backs.

In Fig. 3d, the PET films were firstly corona charged with $V_g$ ranging from 1 kV to 3 kV, obtaining $\Delta V$ of 1 kV to 3 kV. Then the bottom side of all films were CE treated with pure DI water to obtain $V_{fb}$ of -1 kV, not influenced by $\Delta V$. Then, the surface potential of the top side $V_{ft}$ were confirmed to be equal to $V_{fb}+\Delta V$. As all samples shared the same $V_{fb}=$ -1 kV, $V_{ft}$ was manipulated by varying $\Delta V$. It means that, for an insulating polymer film, CD can create a long-lasting $\Delta V$, if one side of it is treated with CE to obtain a given potential, the potential on the other side can be controlled by adding the potential difference $\Delta V$. We further validated this phenomenon by the results in Fig. 3e. All the PET films were CD treated to obtain $\Delta V=1$ kV in the first step. Their bottom side were then treated with IPA solutions of different concentrations. With $V_{fb}$ ranging from -1 kV to 0 by increasing the concentration of IPA, $V_{ft}$ increased from 0 to 1 kV in parallel with $V_{fb}$, with $\Delta V$ as a consistent difference. Thus, the surface voltage of both sides of the films can be well controlled by combining the effect of CD and CE together. Among which, CD controls $\Delta V$, and CE controls the absolute value of $V_{ft}$ or $V_{fb}$, then the potential of the other side can be determined.

## 3.4 Self-stabilization of Surface Potential

As CE treatment can affect the potential value of the films significantly by changing a small amount of the surface free charges. It is reasonable to concern that with time going on or by random touching, the CE induced surface potential may decay to 0 or lose control, ending up with the collapse of the EPAM structure.

To test the stability of the surface potential provided by CD+CE treatment, the surface potential evolution of the CD+CE treated samples were monitored for 7 days. In Fig. 3f, PET films were firstly CD charged with various $\Delta V$=1 kV, 1.5 kV and 2 kV. Then, all films went through CE treatment with pure IPA applied at their top side to obtain the same $V_{ft}$=0. At the beginning, all films obtained $V_{fb}$ of -1 kV, -1.5 kV and -2 kV, as $V_{fb}= V_{ft} - \Delta V$. During the stability test, both $V_{fb}$ and $V_{ft}$ moved upwards in Fig. 3f, meaning $V_{fb}$ and $V_{ft}$ tend to reach the same absolute value till $|V_{ft}|=|V_{fb}|=|0.5\Delta V|$. In Fig. 3g, samples with the same $\Delta V$=1.5 kV, but different initial $V_{fb}$ were tested as well. IPA aqueous solution of different concentrations were utilized to obtain initial $V_{fb}$ of 0, -0.5 kV and -1 kV. After 4 days, $|V_{ft}|=|V_{fb}|=|0.5\Delta V|$ was also obtained. Therefore, no matter how much the original surface potential was, the absolute value of each side of the films tend to spontaneously become $0.5\Delta V$ and $-0.5\Delta V$ on each side. It can be denoted as a spontaneous balancing behavior. Because of its spontaneous recovery capability, it can allow us to maintain the surface potential against disturbances, such as random touch or liquid wiping or cleaning as window layers. This interesting phenomenon can help us to obtain controllable and stable surface potential on both sides of the polymer films, and then maintain stable repulsion force between films in EPAM.

Here, it needs to be clarified that for polymer films, surface potential detected by the fieldmeter (Kelvin probe 2.5 cm away from the film) is generated by both the injected charges from CD and free surface charges controlled by CE. Capability in controlling the surface potential, also means the capability to control the density of both types of charges. The size of the film is another factor determining the measured surface potential. As calculated in Fig. 3h, with the increase of the film size, the measured surface potential also increases. It is because for a finite sized film, more charges will contribute to the measured surface potential. However, the increase slows down with further increase of the film size, as the charges far away from the detect point will contribute less. To eliminate the influence of film size, we will use charge density and repulsive pressure in the following discussion.

**3.5 Electrostatic Repulsion Pressure between Suspended Films**

To maintain stable air gaps in EPAM, the next step is to understand the relationship between charge density and repulsion force between films. Data points in Fig. 4a measured by the set up in Fig. 1f show the molded and measured repulsive pressure between two square shape PET films of 3 inch in length increased with the increase of the surface charge density and decrease of the gap distance between two parallel films. Especially, with distance reducing further, the increase of the repulsive pressure increases dramatically.

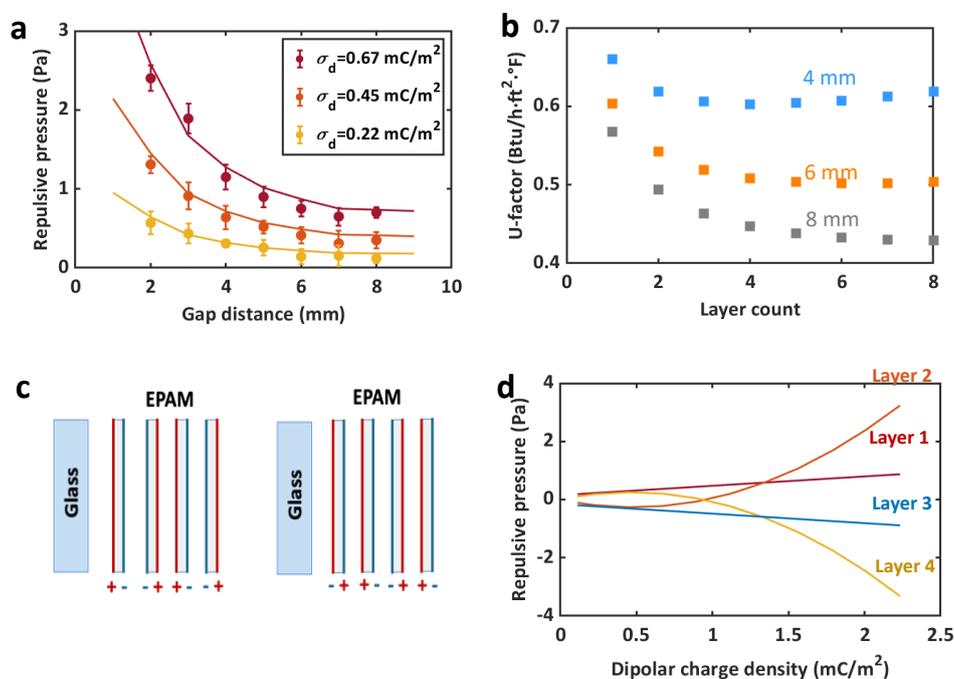

**Fig. 4** Performance of EPAM. a) Repulsive pressure affected by gap distance between two films (3-inch square shape PET film) and charge density, b) U-factor of EPAMs with different total thicknesses and layer count, c) schematic of the polarity distribution for each layer in 4-layer EPAMs, d) influence of charge density on the pressure applied on each layer in a 4-layer EPAM.

According to the modelling results, we can calculate the charge density/surface potential required to generate a given repulsion pressure. Based on the research above, as the surface potential on each side of the film will finalize at $0.5\Delta V$ and $-0.5\Delta V$, we can choose the right $V_g$ to obtain the required $\Delta V$, and then select the concentration of the IPA or NaCl solution to CE treat one side of the film to directly reach the most stable $0.5\Delta V$ and $-0.5\Delta V$ status to create stabilized EPAM.

## 3.6 EPAMs with Different Layers and Thicknesses

Through above disucssion, we learnt that we can provide desirable repulsion force by controlling the long-lasting self-recoverable charges with CD+CE. The last step is to figure out how much pressure is required to create effective EPAM structures to reach the thermal insulating property to reduce energy consumption of windows significantly.

U-factor of EPAM is determined by its thickness and layer count. As shown in Fig. 4b, by increasing the total thickness, the U-factor can be reduced. Increasing the layer count also decreases the U-factor. But no further reduction can be obtained after reaching about 4 layers, because of the increased polymer to air gap thickness ratio. Therefore, we suggest 4-layer EPAM to maximize thermal insulating efficiency at the lowest cost. To create a stable 4-layer EPAM, the polarities of each layer need to be designed as schematized in Fig. 4c. Adjacent side should possess the polarities. Fig. 4d is the influence of charge density on the pressure applied on each layer in a 4-layer EPAM. It can be learnt that the charge density should not be the same on each layer to obtain balanced repulsive force on each layer in the 4-layer EPAM. To stabilize it, the charge density on the outside two layers should be 6 times of the inner two layers, which can be realized by CD+CE as discussed above.

## 3.7 Performance of EPAMs

With CD+CE method, we manufactured 4-layer EPAM with surface potential on each surface as ±2 kV, ±0.5 kV, ±0.5 kV, and ±2kV. The optical image of a 1 ft$^2$ 4-layer EPAMs on a single pane glass is shown in Fig. 5a. It is clear that with the EPAM structure attached onto the glass, no significant Vt reduction or colour change was observed. The Vt measurement results of EPAMs with different layers is elaborated in Fig. 5b. The Vt of PET film is not affected by the charging process, which is still as high as 90.7%. The average Vt values are 83.6% and 70.7% for 2-layer EPAM and 4-layer EPAM respectively.

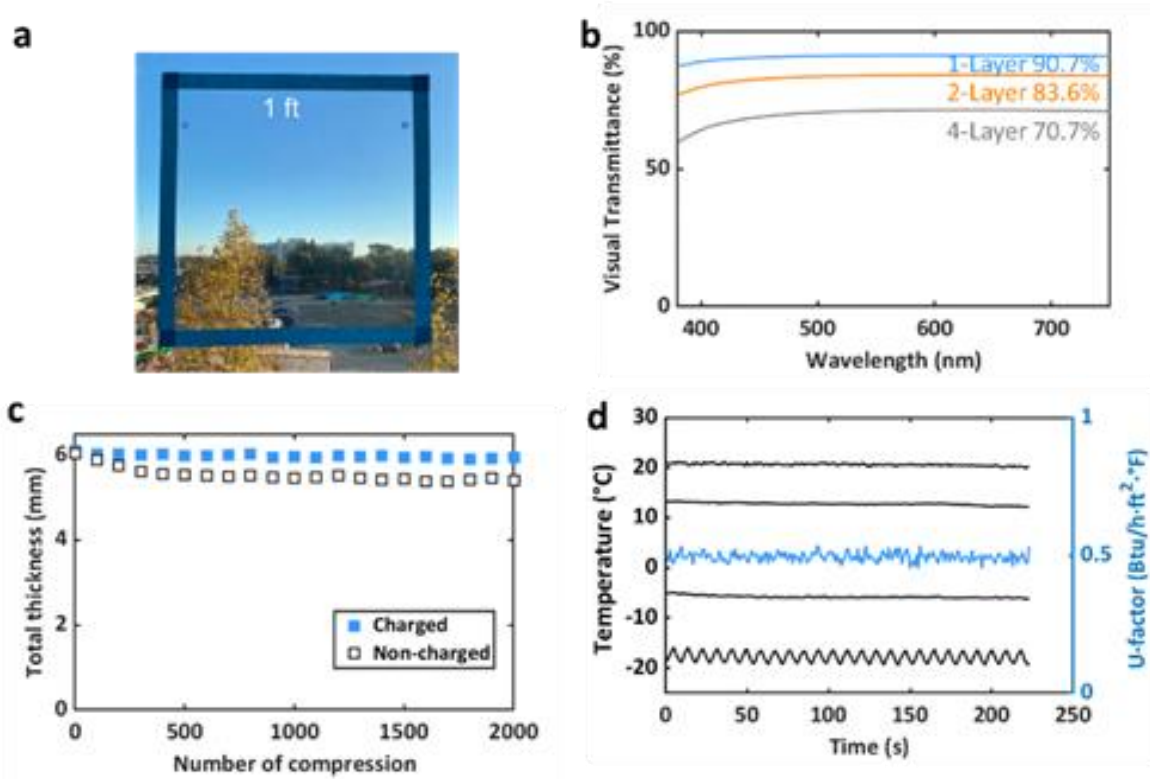

**Fig. 5** Performance of EPAMs. a) Optical images of 1 ft$^2$ 4-layer EPAM on single pane window, b) visual transmittance (Vt) of 1-layer, 2-layer and 4-layer EPAMs at different wavelengths, c) the change of the total thickness of a 6-mm-thick 4-layer EPAM and PAM with the increase of the number of compressions, d) temperatures across 6-mm thick 4-layer EPAM and its measured U-factor; the lines from top to bottom are the indoor temperature, the interior and exterior surface temperature of EPAM and the outside temperature. Blue curve shows the U-factor variation with time.

To compare the structural resilience of PAMs and EPAMs quantitively, compression tests were conducted on them. It turns out that, for a 4-layer EPAM, 2000 times of compression didn't lead to much drop in the total thickness. However, the total thickness of the PAM structure which didn't go through the CD+CE process showed a thickness reduction over 20%. The structural resilience of un-charge PAM is mostly from the support of the frame in the 1-ft-size sample. Larger samples will face more severe collapse. And the thickness reduction will be more severe with the increase of the compression times. Therefore, CD+CE process can significantly increase the structural resilience of EPAM structures.

The thermal insulating property of EPAM was tested with the set up in Fig. 1g. The temperature at the two sides of the EPAM were 13 °C and -7 °C respectively, with temperature difference as large as 20 °C. It is much larger than the temperature difference between a single pane window of as low as 2.5 °C. The U-factor of the 4-layer EPAM structure is as low as 0.5 Btu/hft$^2$°F, comparable with Energy Start windows.

Table 1 Properties of single pane and double pane windows vs. EPAM structures with different layers targeting on U-factor of ~0.5 Btu/hft$^2$°F

| Properties | Single pane | Double pane (clear) [64–66] | 1-layer EPAM | 2-layer EPAM | 4-layer EPAM |
|---|---|---|---|---|---|
| Thickness with glass (mm) | 5 glass | 10 glass + 10~15 gas | 12 PAM +5 glass | 8 PAM +5 glass | 6 EPAM +5 glass |
| Calculated U-factor w/windowpane (Btu/hft$^2$°F) | 1.2 | 0.526 | 0.526 | 0.494 | 0.508 |
| Thermal conductivity (W/mK) | 0.58~1.9 | 0.563 | 0.028±0.02 | 0.029±0.02 | 0.030±0.02 |
| Visual transmittance (%) | 88~90 | 45~80 | 90.7 | 83.6 | 70.7 |
| Color rendering index (CRI) | / | >98 | 99.6 | 95.8 | 95.6 |
| Haze (%) | 0.1 | 0.2 | 0.2 | 0.4 | 1.6 |
| Price ($/ft$^2$) | / | >50 | ~5 | ~6.1 | ~7.4 |

Table 1 compares the performance of single pane, double pane, with 1-layer, 2-layer and 4-layer EPAMs. The targeted U-factor for EPAMs is set as around 0.5 Btu/hft$^2$°F. To reach this U-factor, as discussed in Fig. 4b, the required EPAM thickness decreases with the increase of the layer numbers, which is 12 mm for 1-layer EPAM, 8 mm for 2-layer EPAM and 6 mm for 4-layer EPAM. Meaning, 4-layer EPAM retrofitted single-pane windows can provide similar U-factor as double pane windows with less than half of its total thickness. The thermal conductivity increased slightly as more highly thermal insulating air (0.024~0.026 W/m·K) is replaced by less-insulating PET film (0.15~0.3 W/m·K). The increased radiation and reduced convection help to reach lower U-factor. Plus, Vt reduces with the increase of the layer numbers, while haze increases. In the range of one to four layers, end users can choose their EPAM structure by evaluating their preferences in Vt, thickness and U-factor.

The price of EPAM was calculated by considering labor, material, and the tool kit. It turns out that the cost to retrofit windows with EPAMs is lower than $7.4 $/ft$^2$, at least 10 times cheaper than replacing single pane windows by double pane windows. The reduced heat flux through window can be evaluated by

$$q = (U_{\text{original}} - U_{\text{ELEA}})\Delta T$$

where, $\Delta T$ is the temperature difference through the window pane.[56] If 6-mm-thick 4-layer EPAM is deployed on window glass, the saved heat energy of EPAM retrofitted single pane windows will be on the order of $10^6$ kJ/m$^2$ per year for the US. Thus, EPAM is a cost and energy efficient solution for single pane window retrofitting. If broadly deployed, billions of energy cost will be saved for the US. If broadly used, EPAM may attract massive attention in the window retrofitting market because of its outstanding thermal, Vt and price performances.

**4. Conclusions**

In this paper, electrostatic force suspended polymer-air multilayer structure (EPAM) was developed to provide a high visual transmittance (Vt), low haze and low cost solution to retrofit single pane windows and reduce the massive energy lost through them. Electrostatic force is proved to be able to stabilize the suspended large size flexible polymer films without deteriorating the Vt of the window+. Corona discharge (CD) is able to provide long-lasting charges for insulating polymer films, but it can only control the surface potential difference between two surfaces of the films, not the potential on each side. To guarantee the successful assembly of the EPAM structures, liquid-solid contact electrification (LS-CE) was confirmed to possess the capability to control the potential value of surface potential of non-conductive films by tailoring the ingredient and/or concentration of the low cost and environmentally friendly aqueous CE solutions. By combining CD and CE, the capability to precisely control the surface potential of both sides of transparent non-conductive films can be obtained. The spontaneous balancing behavior of CD+CE charged films can maintain the stability of the surface potential of the films, improving tolerance against random touch or potential decay along time. Controllable surface potential provides controllable electrostatic force to create structural resilient EPAM structures. Single pane windows retrofitted by 4-layer 6-mm-thick EPAMs can provide U-factor of ~0.5 Btu/hft$^2$°F, Vt higher than 70%, haze lower than 1.6%, and color rendering index higher than 95%. Billions of energy cost can be saved if EPAM can be broadly deployed at cost of at least 10 times lower than double pane windows. And combining CD and CE provides an efficient way to realize precise surface potential control. It can be utilized in other applications such as increasing the efficiency of triboelectrification energy harvesters. It will also inspire the innovation of more types of cost and efficient ways for energy saving.

**Conflicts of interest**

There are no conflicts to declare.

**Acknowledgements**

This research was support by the Advanced Research Projects Agency-Energy (ARPA-E) under Grant No. DE-AR0000737 and the start-up funding for Ying Zhong by University of South Florida.**References**